\documentclass[article,amsmath,amssymb]{revtex4}

\usepackage{graphicx}
\usepackage{dcolumn}
\usepackage{bm}
\usepackage{amsmath,amssymb}

\begin{document}


\title{A genetic algorithm to build diatomic potentials}
\author{Luiz Fernando Roncaratti}
\altaffiliation{roncaratti@fis.unb.br}
\author{Ricardo Gargano}
\author{Geraldo Magela e Silva}
\affiliation{Institute of Physics, University of Brasilia}

\date{\today}

\begin{abstract}
Through the last years, several types of numerical and combinatorial optimization
algorithms have been used as useful tools to minimize functional forms. Generally, when
those forms are non-linear or occur in problems without a specific optimization method,
stochastic methods based on search algorithms have shown good results due to its smaller
susceptibility to be trapped in a local minimum. Besides that, they can easily be
implemented to work with other techniques, in this class of algorithms, the genetic ones
have received special attention because they are a robust optimization tool. An algorithm
can be named genetic when it uses some kind of codification to transform a set of possible
solutions of a given problem in a population that will evolve subject to operators
inspired, or not, by mechanisms of natural selection. In other words, they work with a
population of solutions to obtain better solutions in the next generation. To do this, they use
only information of cost and prize. In this work, we propose a genetic algorithm
optimization technique (GAOT) to fit diatomic potential energy curves.
In order to show this method, we obtain the analytical functions of the $H_{2}$$^{+}$ and $Li$$_2$
systems using the $ab$ $initio$ energy calculations as well as Rydberg
trial function. These studies show that the quality of the GAOT fitting is comparable
to the best optimization techniques recommended to fit diatomic systems. The introduction
of this new technique is very important because it arises as a new option to fit potential
energy surfaces for reactive scattering dynamics.
\end{abstract}

\maketitle

\section{Introduction}

Genetic algorithms \cite{genetico1,genetico2,genetico3, genetico4, genetico5, genetico6} have been applied
successfully in the description of a variety of global
minimization problems as well as have attracted significant attention due to their suitability for large-scale
optimization problems, specially for those in which a desired global minimum is hidden among many local minima.

The main object of this paper is to propose a genetic algorithm technique for fitting the potential energy
curves (PEC) to points obtained by {\it ab-initio} calculations. In order to present and to test the method,
we reproduce the PEC of the following diatomic systems $H_{2}$$^{+}$ and $Li$$_{2}$ using a Rydberg functional form \cite{varandas}.
The molecular constants and PEC of the and lithium dimer have been of great interest to theoretical chemists, spectroscopists and
astrophysicists. The $H_{2}$$^{+}$ system is the most simple and probably the most studied molecule.
These molecules are relatively small and can be treated very accurately. It is not a surprise that they are used
for testing and applications of new methodologies \cite{gargano}. Further, these systems has been intensely
investigated in experimental basis.

Following, we outline how the paper is organized. In Section 2 we present the main characteristics of the GAOT
used in our calculations. The details of the fitting are shown in Section 3.
Our conclusions and detailed comparison with other methods are contained in the Section 4.

\section{Model}

\subsection{The Problem}

In order to find the PEC of $H_{2}$$^{+}$ and $Li$$_2$ we want that the GAOT finds a set of parameters
$[\bold{a}]=[a_1,a_2,..,a_m]$ that minimize the mean square deviation of Rydberg functional form. We employed
this functional form to adjust to {\it ab initio} points of these diatomic systems. Given a set of $n_p$ ordered
pairs $(e_{p},r_{p})$ of {\it ab initio} points, where $e_{p}$ is the energies and $r_{p}$ is the distance
between the nuclei, we want to find a functional form $V([\bold{a}],r)$ that minimizes the mean square deviation
\begin{eqnarray}
S=\sum_{p}^{n_p}\delta_p^2=\sum_{p}^{n_p}(e_{p}-\overline{e}_{p})^2
\end{eqnarray}
where $\overline{e}_{p} \equiv V([\bold{a}],r_{p})$.

In this work we employ the Rydberg form, $V^{ry}([\bold {a}],\rho)$, to adjust the potential of diatomic systems
through a stochastic optimization technique based on GAOT. Our method is general and we can calculate the PES of
systems with more degrees of freedom. To find the potential form of a given system, we need a set of points that
we want to adjust. With the total energies as a function of the nuclear configurations we have to choose an
appropriate compact form to represent them. In this work, the Rydberg form is given by
\begin{equation}
V^{ry}([\bold {a}],\rho)=-D_{e}(1+\sum_{j=1}^{m^{'}}a_{j}{\rho}^j)e^{-a_{1}{\rho}}
\end{equation}
The dissociation energy is given by
\begin{equation}
D_{e}=-V^{ry}([\bold {a}],0)
\end{equation}
where $\rho=r-r_{eq}$ and $r_{eq}$ is the equilibrium bond length of the system.
In this work we use $m^{'}=3$ for the Rydberg form.

\subsection{Codification}

In our GAOT the population is coded in a binary discrete cube named $\bold{A}$, with $l\times m\times n$ bits.
The elements of $\bold{A}$, $a_{ijk}$, are either 0 or 1, with $i,j,k$ integers numbers $1\leq i\leq n$,
$1\leq j \leq m$, $1\leq k \leq n$. The label $i$
refers to the component $i$ of the {\it gene} $j$ of the individual $k$. Therefore, $A$ represents a
population of $n$ individuals, each one of them have a genetic code with $m$ {\it genes}. Each {\it gene} is a
binary string with $l$ bits.

The genetic code of the individual $k$ is given by
\begin{equation}
[\overline{\bold{a}}]_k=[\overline{a}_{1k},\overline{a}_{2k},...,\overline{a}_{mk}],
\end{equation}
\begin{equation}
\overline{a}_{jk}=\sum_{i=1}^{l}2^{i-1}a_{ijk}
\end{equation}
is a integer number composed with the binary string $a_{1jk}a_{2jk}..a_{ijk}..a_{ljk}$. It is defined on the
interval $[0,2^{l}-1]$. To define the real search space for each parameter, we transform
\begin{equation}
\overline{a}_{jk} \rightarrow {a}_{jk} \equiv
\frac{(a^{max}_{j}-a^{min}_{j})}{2^{l}-1}\overline{a}_{jk}+a^{min}_{j}
\end{equation}
were ${a}_{jk}$ is a real number defined on the interval
\begin{equation}
\delta_j=[a^{min}_{j},a^{max}_{j}].
\end{equation}
Now we define the phenotype of the individual $k$
\begin{equation}
V^{ry}_{kp} \equiv V([{\bold{a}}]_k,\rho_{kp}) =
-a_{5k}(1+a_{1k}{{\rho}_{kp}}+a_{2k}{{\rho}_{kp}}^2+a_{3k}{{\rho}_{kp}}^3)e^{-a_{1}{\rho}_{kp}}
\end{equation}
where $[\bold{a}]_k=[a_{1k},a_{2k},a_{3k},a_{4k},a_{5k}]$, ${\rho}_{kp}=r_{p}-a_{4k}$ and $r_{p}$ is the
interatomic distance of the system. The gene $a_{5k}$ represents the dissociation energy. By this way we have
$m=m^{'}+2=5$ independent parameters to optimize. With this we define the fitness of a phenotype $k$
\begin{equation}
F_k=S_{max}-S_k
\end{equation}
\begin{equation}
S_k=\sum_{p}^{n_p}(\delta_{kp})^2=\sum_{p}^{n_p}(e_{p}-V^{ry}_{kp})^2
\end{equation}
where $S_{max}$ is worst individual in the population.

\subsection{Operators}

In our GAOT we use the most common operators: selection, recombination and mutation. The selection operator
normalize the vector $F_{k}$
\begin{equation}
P_{k}={\frac {F_{k}}{\sum F_{k}}}
\end{equation}
that represents the probability of each individual been selected for a recombination through a roulette
spinning. For the purpose of this work we selected $n/2$ individuals (parents) that will generate, through the
recombination operator, $n/2$ new individuals (offsprings). So, to make a new generation we joint the $n/2$
old strings (parents) with a $n/2$ new strings (offsprings) in order to maintain a population with fixed
number $n$. The recombination
operator is a cross-over operator that recombine the binary string of each gene $j$ of two random selected
individuals to form two new individuals. In this work we use a two random point cross-over.

The mutation operator flip $N_{mut}$ random selected bits in a population. We choose $N_{mut}$ to make the
probability of change of a given bit equal to $0.01$ per cent. So, in a population of $l\times m\times n$ bits, we
make
\begin{equation}
 q = \frac {N_{mut}}{l\times m\times n}
\end{equation}
where $q$ is the probability of change in one bit.

\subsection{Linear scaling and elitist strategy}

When we use the GAOT in a minimization procedure we want to find a solution that is a global minimum. In fact,
this solution can be found by a group of individuals or by all individuals in a population. When all the
population converges to a single solution, this solution could not be a global minimum, but a local minimum.
This is called premature convergence and it can be avoided by the linear scaling \cite{genetico1}. This
procedure enhances the probability that several minima will coexists in the population. A general linear scaling
in $S_{k}$ is given by

\begin{equation}
S^{'}_{k} = aS_{k}+b.
\end{equation}
In this work we define
\begin{equation}
e=\frac {S_{max}-g\overline{S}}{g-1}
\end{equation}
if $S_{min} \geqslant e$
\begin{eqnarray}
a&=&\frac {(g-1)\overline{S}}{S_{max}-\overline{S}}\nonumber\\
b&=&\frac {(S_{max}-g\overline{S})\overline{S}}{S_{max}-\overline{S}}
\end{eqnarray}
and if $S_{min}<e$
\begin{eqnarray}
a&=&\frac {\overline{S}}{\overline{S}-S_{min}}\nonumber\\
b&=&-\frac {\overline{S}S_{min}}{\overline{S}-S_{min}}.
\end{eqnarray}

Where $S_{max}$ and $S_{min}$ are the worst and the best phenotypes, respectively. $\overline{S}$ is the mean
value of the phenotypes of the population. In this way, we maintain the average
$\overline{S'_{k}}=\overline{S_{k}}$, set ${S'}_{max}=g\overline{S}$ if ${S}_{max} \gg \overline{S}$ and
${S'}_{min}=0$ if ${S}_{min} \ll \overline{S}$. $g$ is a arbitrary value to control the selective pressure.
Expression (14) avoids negative values for ${S'}_{k}$. The selection operator normalize the vector
$F^{'}_{k}=g\overline{S}-S^{'}_{k}$. In fact, this linear scaling has lowered the selective pressure on the
population. Through this, and setting a correct mutation rate, we maintain the variety of the population and
therefore we avoid the convergence of the population. The elitist strategy consists of copying an arbitrary
number $N_{el}$ of the best individual on the population in the next generation. It warrants that this
individual will not be extinguished. Here a example is useful. We want find the minimum of the one dimensional
test function
\begin{eqnarray}
V(x)=-(&&20e^{-0.2(x-7.5)^2}\,+\,\,2e^{-0.09(x-15)^2}+\nonumber\\
       &&\,\,5e^{-0.09(x-20)^2}\,+12e^{-0.09(x-28)^2}+\nonumber\\
       &&22e^{-0.09(x-35)^2}+\,\,2e^{-0.09(x-50)^2}+\nonumber\\
       &&\,\,7e^{-0.2(x-42.5)^2}\,\,)\nonumber
\end{eqnarray}
at the interval $[0,50]$. In Fig. 1 we show the form of this function and the frequency of each individual
appearing in the population, after 250 and 500 generations. We present the results with $S'_{k}$ or $S_{k}$.
It is clear the effect
of the linear fit in the population dynamics. When we use $S'_{k}$, after 500 generations, the 880 individuals in
the population make a signature of the function V(x).

\begin{figure}[t]
\includegraphics{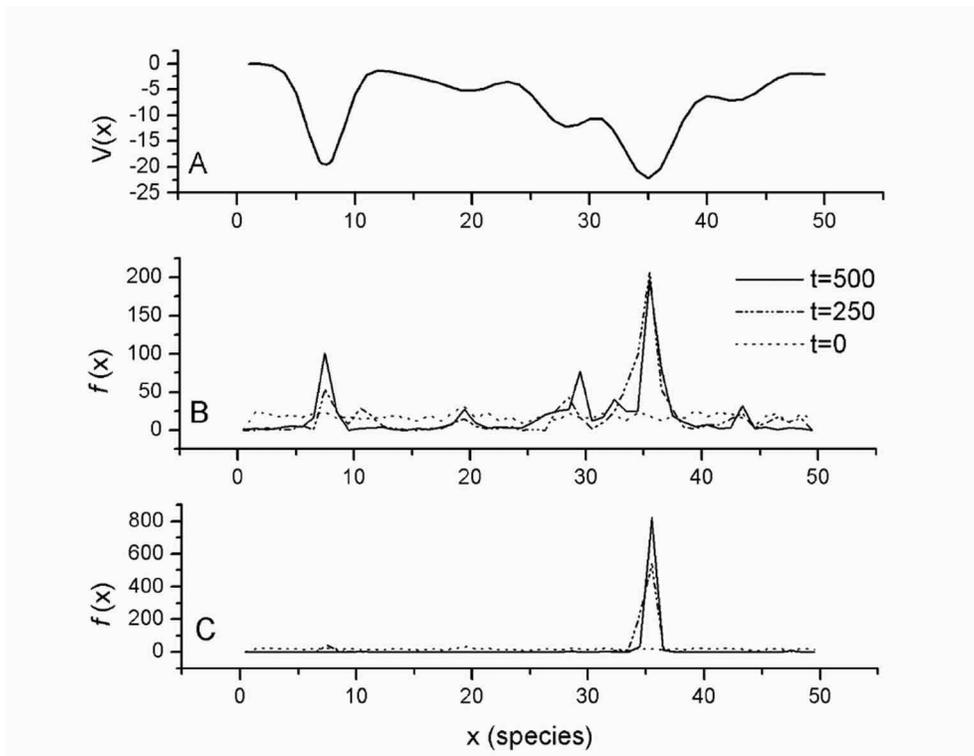}
\caption{(A) Test function V(x). Frequency of each individual in the population after 250 and 500 generations
with (B) linear fit on (g=1.1,$N_{el}=8$) and (C) linear fit off.}
\label{figure1}
\end{figure}

\section{Results}

We run 50 times the algorithm for each case to evaluate the performance of the method.
After extensive trials of the parameters values we take as parameters for all cases:
$n$=800, $l$=30, $m$=6 or 8, $q$=0.01, $N_{el}$=40 and a 1000 generations. It should
be pointed out that the algorithm is very robust and it works properly with an wide
range of these parameters. The initial population were always random numbers.

All the runs produced a subpopulation of acceptable solutions. We defined as acceptable solutions when the root
mean square (rms) and the difference per point are smaller than $1.0$ Kcal/mol. Actually, for each run, we found
a great number of acceptable solutions. Figure 2 shows the distribution of individuals in the region of interest
(acceptable solutions) after a 1000 generation of the fifty runs for Li2 PEC. Similar results has been found for
H2+.

\begin{figure}[h]
\includegraphics{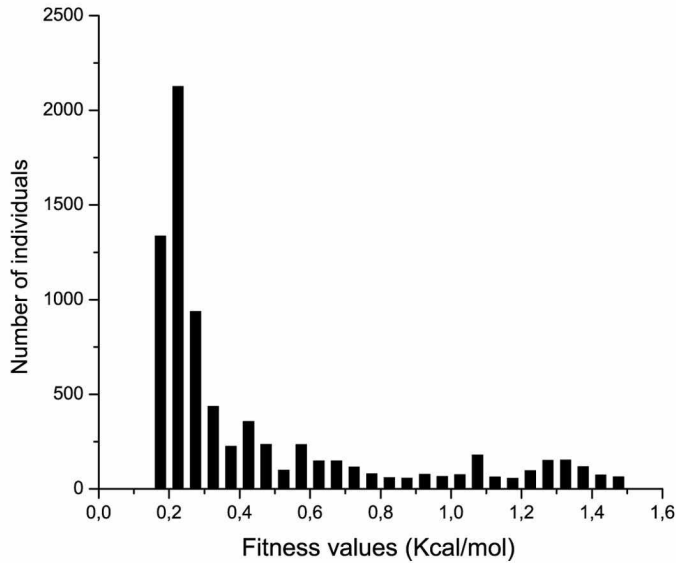}
\caption{Distribution of individuals in the region of interest (acceptable solutions) after a 1000 generation of
the fifty runs for Li2 PEC.} \label{figure1}
\end{figure}

\begin{figure}[!]
\includegraphics{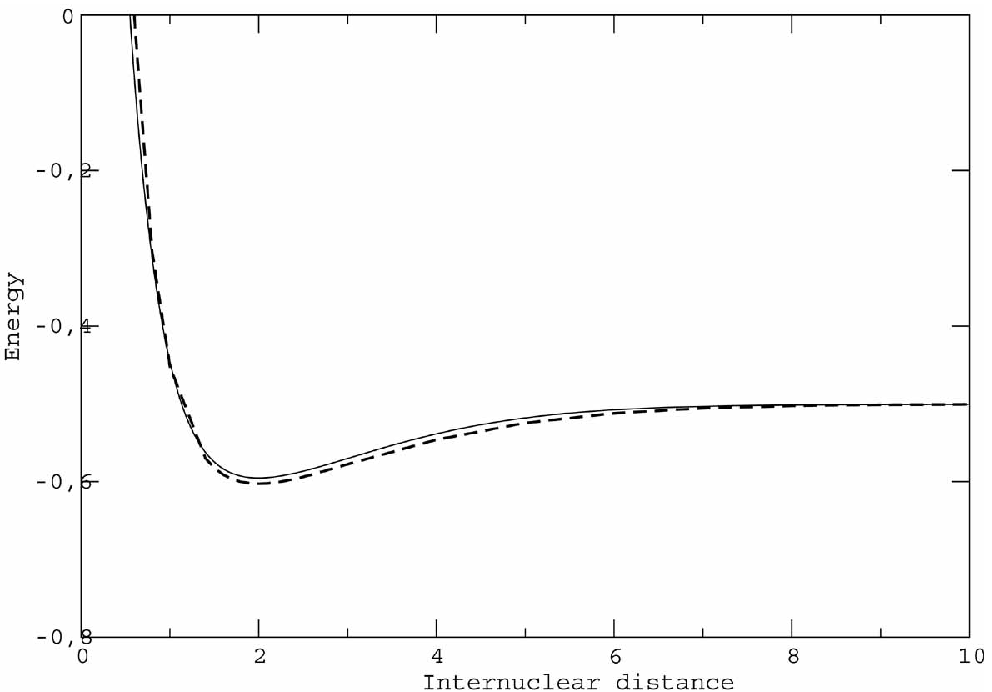}
\caption{Comparison between the GAOT(solid line) and {\it ab initio}(dashed line) PEC of $H_{2}$$^{+}$ system.}
\label{figure2}
\end{figure}

Initially we employed the GAOT to obtained the potential curves of $H_{2}$$^{+}$ ion molecule using a set of
highly accurate molecular energies (table 1 of reference \cite{articleh2+}) and Rydberg function (see equation
2) with $r_{eq}$ and $D_{e}$ free to change. In the Rydberg form $r_{eq}$ and $D_{e}$ are the {\it genes}
$a_{4}$ and $a_{5}$ respectively. So, we chose their search intervals, $\delta_4$ and $\delta_5$, close to the
regions were we have experimental values for these parameters \cite{articleh2+}. In Table \ref{table1} are shown
the parameters obtained by our fitting using the GAOT. The rms deviation value $S$ in this fitting was of the
$0.927\times 10^{-04}$ Hartree (about $0.05810$ kcal/mol). The figure \ref{figure2} shows a comparison between
the GAOT and {\it ab initio} PEC. From this figure one can see that both GAOT and {\it ab initio} PEC are in a
good agreement. Once a PEC has been fitted to a analytical form, both diatomic vibrational energies and spectra
can be determined from the radial Schrodinger equation \cite{nuclear}. So, to better test our fitting, we
resolved the radial Schr\"odinger equation using our $H_{2}$$^{+}$ PEC through DVR method \cite{dvr}. The Table
\ref{table2} shows the comparison among the spectra values obtained using the GAOT PEC with the spectra of the
reference \cite{articleh2+}. One can see from these comparison that GAOT spectra are in good agreement with that
found in the literature.

\begin{figure}[!]
\includegraphics{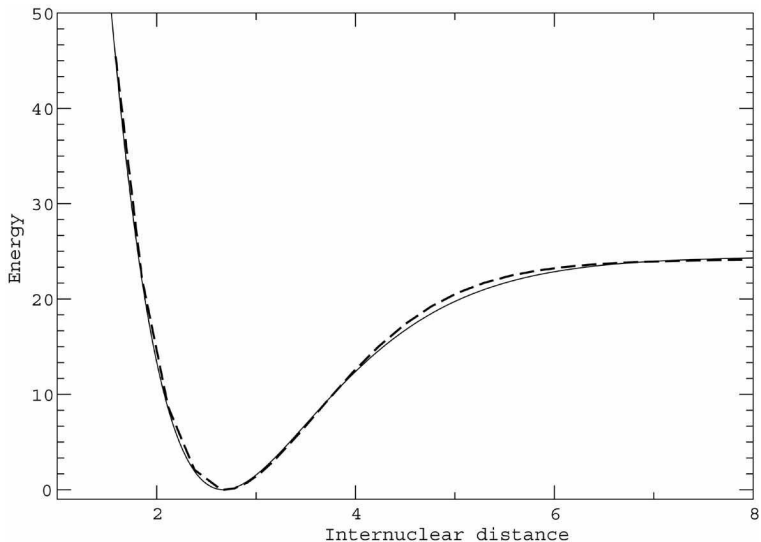}
\caption{Comparison between the GAOT(solid line) and {\it ab initio}(dashed line) PEC of $Li$$_2$ system.}
\label{figure3}
\end{figure}

To complete our test about of efficiency of the GOAT, we also fit the PEC for the $Li$$_2$ system from {\it ab
initio} energies. To make this we used the {\it ab initio} energies of reference \cite{paulo} and the Rydberg
analytical function. We obtained good results to this molecule with $r_{eq}$ and $D_{e}$ free parameters in the
minimization process. In this case, the rms deviation value was about $0.3062884602$ kcal/mol, $r_{eq}=2.6732799
A^{o}$ and $D_{e}=24.44413$ kcal/mol.  The GAOT results obtained for $Li$$_2$ molecule are in Table
\ref{table3}. In figure \ref{figure3} are represented both GAOT and {\it ab initio} PEC of the $Li$$_2$ system.
From this figure one can see that GAOT and {\it ab initio} PEC are in a good agreement. In the Table
\ref{table4}, we compare the $Li$$_2$ spectra obtained via GAOT PEC and two others spectra found in the
literature, FCIPP \cite{paulo} and RKR \cite{paulo}, respectively. From this comparison one can note that GAOT
spectra are in good agrement with both FCIPP and RKR spectra.

\def\tablename{Table}
\begin{table}[h]
\caption{Parameters obtained of GAOT fitting for $H_{2}$$^{+}$ in Rydberg form. Energy and nuclear distances are given in  atomic units.}
\centering
\begin{tabular}{|l|l|}
\hline \,Parameters\,          &\,$H_2$$^{+}$\,
\\\hline\hline \,$a_{1}$\,     &\,1.4960937607\,
\\\hline \,$a_{2}$\,           &\,0.5767577799\,
\\\hline \,$a_{3}$\,           &\,0.21384215222\,
\\\hline \,$r_{eq}$\,          &\,1.9970569317\,
\\\hline \,$D_{e}$\,           &\,0.0946586828\,
\\\hline
\end{tabular}
\label{table1}
\end{table}

\def\tablename{Table}
\begin{table}[h]
\caption{$H_2$$^{+}$ spectra given in $cm$$^{-1}$} \centering
\begin{tabular}{|l|l|l|}
\hline \,\,           &\,GAOT spectra\,     &\,Reference \cite{articleh2+}\,
\\\hline\hline \,1\,     &\,2190.1104\,       &\,2192.022\,
\\\hline \,2\,           &\,4256.9176\,       &\,4256.714\,
\\\hline \,3\,           &\,6204.1122\,       &\,6198.318\,
\\\hline
\end{tabular}
\label{table2}
\end{table}

\def\tablename{Table}
\begin{table}[h]
\caption{Parameters obtained of GAOT fitting for $Li$$_2$ in a Rydberg form. Energy in Kcal/mol and nuclear distances in Angstrons.}
\centering
\begin{tabular}{|l|l|}
\hline \,Parameters\,           &\,$Li$$_2$\,
\\\hline\hline \,$a_{1}$\,     &\,1.91967773\,
\\\hline \,$a_{2}$\,           &\,1.078125\,
\\\hline \,$a_{3}$\,           &\,0.22248840\,
\\\hline \,$r_{eq}$\,          &\,2.69008356\,
\\\hline \,$D_{e}$\,           &\,24.4238281\,
\\\hline \,$S$\,               &\,0.11981447\,
\\\hline
\end{tabular}
\label{table3}
\end{table}
\def\tablename{Table}

\def\tablename{Table}
\begin{table}[h]
\caption{$Li$$_2$ spectra given in $cm$$^{-1}$}
\centering
\begin{tabular}{|l|l|l|l|}
\hline \,\,           &\,GAOT spectra\,     &\,FCIPP\cite{paulo}\,     &\,RKR\cite{paulo}\,
\\\hline\hline \,1\,     &\,347.42\,        &\,346.05\,             &\,346.46\,
\\\hline \,2\,           &\,688.69\,        &\,686.65\,             &\,687.86\,
\\\hline \,3\,           &\,1023.79\,       &\,1021.71\,            &\,1024.08\,
\\\hline \,4\,           &\,1352.65\,       &\,1351.15\,            &\,1355.01\,
\\\hline \,5\,           &\,1675.21\,       &\,1674.88\,            &\,1680.54\,
\\\hline \,6\,           &\,1991.43\,       &\,1992.81\,            &\,2000.56\,
\\\hline \,7\,           &\,2301.24\,       &\,2304.85\,            &\,2314.95\,
\\\hline \,8\,           &\,2604.59\,       &\,2610.92\,            &\,2623.58\,
\\\hline \,9\,           &\,2901.42\,       &\,2910.90\,            &\,2926.35\,
\\\hline \,10\,          &\,3191.66\,       &\,3204.70\,            &\,3223.11\,
\\\hline
\end{tabular}
\label{table4}
\end{table}

\section{Conclusions}

The genetic algorithm is a useful minimization tool. We can use it to obtain the potential curve of diatomic
systems by adjust {\it ab initio} or experimental data. The dynamics of the population in the GAOT given us
important information about the studied system, in special when we find the distribution of the local minima.
Understand this dynamics and developing tools and theories to control them is important for employed the GAOT in
several classes of physical problems. This study was very important, so it enables the GAOT to fit potential
energy surface (PES) of the bond and scattering systems.

\section{Acknowlegment}
The authors would like to know the Brazilian Science and Technology Council (CNPq), CAPES and FINATEC by support
financial.


\section{References}
\begin{thebibliography}{9}

\bibitem {genetico1} D. A. Goldberg. Genetic Algorithms in Search, Optimization, and Machine Learning. Addison-Wesley
Publishing Company, Inc., 1989.
\bibitem {genetico2} J. H. Holland, Adaptation in natural and artificial systems, MIT press, 1992.
\bibitem {genetico3} D. M. Deaven and K. M. Ho, Phys. Rev. Lett. 75, 288-291 (1995).
\bibitem {genetico4} F. Starrost, S. Bornholdt, C. Solterbeck, and W. Schattke, Phys. Rev. B 53, 12549-12552 (1996).
\bibitem {genetico5} A. Prugel–Bennett and J. L. Shapiro, Phys. Rev. Lett. 72, 1305-1309 (1994) .
\bibitem {genetico6} V. E. Bazterra, O. Oña, M. C. Caputo, M. B. Ferraro, P. Fuentealba, and J. C. Facelli
Phys. Rev. A 69, 053202 (2004).
\bibitem {varandas} J. N. Murrel, S. C. Farantos, P. Huxley, and A. J. C. Varandas, Molecular Potencial Energy Functions,
Wiley Chinchester, 1984.
\bibitem {gargano} W. B. Silva, E. A. Corr\^ea, P. H. Acioli and R. Gargano, International Journal of
Quantum Chemistry. 95 (2003)
149-152.
\bibitem {articleh2+} S. A. Alexander, R. L. Coldwell, Chemical Physics Letters. 413 253-257 (2005).
\bibitem {nuclear} L. Pauling, E. B. Wilson, Introduction to Quantum Mechanics, McGraw-Hill, New York, 1935.
\bibitem {dvr} A. S. Dickinson and P. R. Certain, J. Chem. Phys. 49 (1968) 4209.
\bibitem {paulo}Angelo M. Maniero and Paulo H. Acioli, International Journal of Quantum Chemistry 103, 711-717 (2005).
\end{thebibliography}
\end{document}